# Tuning of tetrahedrality in a silicon potential yields a series of monatomic (metal-like) glassformers of very high fragility.

Valeria Molinero[1]*, Srikanth Sastry[2], and C. Austen Angell[1]

[1]Department of Chemistry and Biochemistry, Arizona State University, Tempe, AZ 85287, U.S.A.

[2]Jawaharlal Nehru Center for Advanced Scientific Research, Jakkur Campus, Bangalore, 560064, India.

We obtained monatomic glassformers in simulations by modifying the tetrahedral character in a silicon potential to explore a triple point zone between potentials favoring diamond (DC) and BCC crystals. DC crystallization is always preceded by a polyamorphic transformation of the liquid, and is frustrated when the Kauzmann temperature of the high temperature liquid intersects the liquid-liquid coexistence line. The glassforming liquids are extraordinarily fragile. Our results suggest that Si and Ge liquids may be vitrified at a pressure close to the diamond-β-tin-liquid triple point.

Humans have made silicate glasses as glazes or in bulk for at least 8,000 years. But it was not until 1960 that the first metallic glass was obtained by cooling a $Au_{75}Si_{25}$ melt at a rate of $q=10^6$ K/s.[1] Metallic alloys that vitrify for $q$ as low as 0.005 K/s[2] have been developed since then, fostering technological applications for bulk metallic glasses. The successful glass formation with alloys notwithstanding, no monatomic glass –metallic or otherwise- has been obtained to date by cooling of a melt. Glass formation in monatomic systems remains a challenge even in computer simulations, where the small system sizes and high cooling rates disfavor crystal nucleation. The monatomic model of Dzugutov[3] can be supercooled considerably, but it ends up forming quasicrystals in isochoric simulations[4] and FCC or BCC crystals in isobaric simulations.[5] Given the lack of true monatomic glassformer models for simulations, the most fundamental studies of the deep supercooled regime in atomic liquids had to be conducted with a binary mixture of Lennard-Jones particles (BMLJ) that mimics the marginal glassformer $Ni_{80}P_{20}$.[6]

In this Letter we develop a strategy for making monatomic glassformers in simulations, and we apply it to the design of a monatomic model that resists crystallization and quasicrystal formation over hundreds of nanoseconds of zero pressure molecular dynamics (MD). We address the key question of *why* the model has slow crystallization kinetics, in terms of the factors that control crystal nucleation. We finally generalize our strategy to propose a way to make the elusive monatomic glassformer in the lab.

Metallic alloys with high glass forming ability (GFA) are multi component systems, and their main constituents have a negative enthalpy of mixing.[7] The latter produces deep eutectics, around which the GFA is highest. In our strategy to develop monatomic glassformers, we start with a poor glassformer, the Stillinger-Weber (SW) model for silicon[8], and vary the interatomic potential –the interactions instead of the composition- to find a deep pseudo-eutectic or low temperature triple point between diamond cubic (DC), body centered cubic (BCC) crystals, and the liquid. The results are described by a novel temperature-potential phase diagram with a glassforming domain around the low temperature triple point.

In the SW model, tetrahedral coordination is favored by adding to a basic pair-wise potential, $v_2(r)$, a three-body term, $v_3(r,\theta)$, which induces repulsion for angles that are not tetrahedral, $v=v_2(r) + \lambda\, v_3(r,\theta)$. The repulsion parameter, $\lambda=21$, and the pair potential parameters were adjusted in [8] to best reproduce the crystalline ground state, density and cohesive energy for the laboratory substance. We have kept the pair potential, defining an invariant temperature scale, and varied λ to tune the repulsive potential in the range 21.5 > λ > 15. The results were obtained from a series of NPT molecular dynamics (MD) simulations for 512 atoms (686 if starting from perfect BCC crystals) at pressure p=0, using procedures given elsewhere.[9] Run lengths ranged from 1 to 130 ns. The average displacement of the atoms was at least 4 atomic diameters during each of the equilibrium runs.

In what follows we discuss the phase diagram as a function of the tetrahedral parameter, λ, and analyze the interplay between the three factors that determine the rate at which crystals form during cooling,[10, 11] and hence the glassforming ability: crystallization driving force, liquid diffusivity and structural similarity between the liquid and crystal phases.

It is well known that the low density amorphous semiconductor phase of silicon a-Si is structurally unrelated to the high density liquid (HDL) metallic phase and cannot be obtained by direct cooling of the melt. Simulations of SW Si have established that these two distinct amorphous phases are related by a first order phase transition.[9] In SW Si, the liquid-liquid (LL) transition occurs between two metastable phases, 650 K below the melting temperature of the DC crystal. We confirm that the SW liquid (λ = 21) does not crystallize at temperatures above the LL transition, and that crystallization to DC occurs from the low temperature (low density) liquid.[9] We observe the same pattern of LL transition at $T_{LL}$ (Fig.1) followed by DC crystallization from the low density liquid (LDL) for all systems with λ > 20.25. For λ < 17, cooling of the liquid results in a sharp crystallization to BCC. Crystallization is signaled by a sharp drop in potential energy and the appearance of characteristic crystal peaks in the radial distribution function.

The DC and BCC crystals that form contain defects and are our starting point for the melting lines determination.[12] We repeat the melting study for the DC (BCC) crystal with decreasing (increasing) λ until the crystal becomes so metastable that it melts almost isoenthalpically to a glass. We assemble these two melting lines into a new type of phase diagram, Fig. 1, in which the potential parameter λ replaces pressure on the horizontal axis of the familiar one-component T-p diagram. The triple point defined by the crossing of BCC and DC melting lines occurs for a tetrahedral parameter is $\lambda^{TP}$=18.75, very close to the λ=18.6 we predict from the T=0 K lattice energies.[13]

Cooling of the monatomic liquid around the triple point, in the range 17.5 < λ < 20.25, does not result in crystallization, but in a continuous transformation to a glass. These glasses reversibly transform into liquids on heating, confirming the absence of crystals or critical nuclei. In classical nucleation theory the activation free energy to form a nucleus is inversely proportional to the square of the crystallization driving force[11], $G^{ex}=G^{liquid}-G^{crystal}$, that increases with supercooling. This quantity has been considered crucial for the glassforming ability of metal alloys, where values as small as 1.5 kJ/mol for $T/T_m$=0.8 typify the best glassforming mixtures.[14]. We computed the excess thermodynamic properties shown in Figure 2 from i) the melting temperatures $T_m$, ii) the melting enthalpy $\Delta H_m$ evaluated as the difference between H of the liquid and perfect crystal, at $T_m$ and iii) the heat capacities $C_p$ (derived from the enthalpies) of the supercooled liquids and perfect crystals.[13] The excess entropies are computed as

$$S^{ex}(T) = \frac{\Delta H_m}{T_m} - \int_{Tm}^{T} \frac{C_p^{liquid} - C_p^{crystal}}{T'} dT', \qquad (1)$$

and the excess free energies by $G^{ex}(T)=H^{ex}(T)-TS^{ex}(T)$. Figure 2c shows $G^{ex}(T/T_m)$ for the supercooled liquids with potentials λ=16 to 20.25. The increase of $G^{ex}$ with supercooling is minimal for the triple point potential. For λ=18.5 $G^{ex}$=1.9 kJ/mol at $T/T_m$=0.8 (Fig 2d), comparable to $G^{ex}$ of metallic glassforming alloys that vitrify for $10^4$ Ks$^{-1}$ cooling rate, such as $Zr_{62}Ni_{38}$.[14]

The diffusivity of the liquid is other critical factor controlling crystallization.[11] In Fig. 3 we examine the diffusivity-temperature relations in the glassforming domain using Arrhenius plots to emphasize the strongly super-Arrhenius (i.e. fragile[15]) character of the diffusivity. We find that the isothermal diffusivity is always maximum for the liquid with $\lambda$=17.5-18. Nevertheless, the diffusivity evaluated on the melting lines reaches a minimum of 0.95 x $10^{-5}$ $cm^2s^{-1}$ at the $\lambda$ value closest to the triple point (inset of Fig. 3). This diffusivity is the same as that of Ni in the marginal laboratory glassformer $Ni_{80}P_{20}$ at the Ni-P eutectic temperature of 1171 K.[16]

From the curvature of the Arrhenius plots of Fig. 2, the parameters for the Vogel-Fulcher-Tammann (VFT) equation

$$D = D_o \exp\left(-\frac{T_o}{F(T-T_o)}\right) \quad (2)$$

may be obtained. The values of zero mobility temperatures $T_o$ and fragility $F$ of Eq.(2) are assessed in the $\lambda$ range 17.5-20.25 in which neither liquid-liquid nor crystallization transitions interfere, and supercooling range is limited only by computer time. We find remarkably high fragilities for all these liquids, $F$ about 0.3[13] (*m* fragility[15] ~ 200). The high fragility of the HDL is not unexpected in view of Adam-Gibbs theory[17] that relates the mobility of the liquid $D = D' \exp(C/S_c T)$ to the number of accessible states, measured by the configurational entropy $S_c$ and a constant $C$ containing a material dependent activation energy. Figure 2b shows the rapid temperature variation of $S^{ex}$ (at low T a good approximation for $S_c$[18]), that arises from the rapidly increasing $C_P$ of the supercooled HDL (Fig 2c and [13]), similar to the one observed for the more fragile metallic glassformers.[14]

The extrapolation of the excess entropies to zero defines the Kauzmann temperatures of the HDL, $S^{ex}(T_K)=0$ [18], which we compare in Fig. 1 with the zero mobility temperature $T_o$ of the VFT fits of the diffusivity data (Eq. 2). The coincidence of $T_o$ and $T_K$, despite the considerable $\lambda$-dependence of each, is remarkable (like that found earlier[19] for different densities in the BMLJ system) and in agreement with the predictions of theories that relate mobility and thermodynamics, such as that of Adam and Gibbs. Both in our study of the modified SW Si potential (mod-SW), and in the case of eutectic alloys, glasses form in the vicinity of the minimum melting point. Close to the triple point crystal nucleation is disfavored because the liquids have i) the lowest diffusivity at $T_m$ (and D rapidly decreases below), and ii) a minimum increase of $G^{ex}$ with supercooling.

The third factor that affects nucleation is the structural similarity between liquid and crystal, signaled by their density difference. At the triple point, BCC is 20% denser than DC, and HDL is just 4% lighter than BCC. A big density gap between crystal and liquid decreases nucleation probability: $\lambda=17$ easily crystallizes to BCC, which is just 2% denser, while $\lambda=20$, 9% denser than DC, has comparable diffusivity, much higher $G^{ex}$ and is a glassformer. The similarity in structure between LDL and DC decreases $H^{ex}$, that correlates with the value of the liquid-solid interfacial tension.[10] The polyamorphic transformation is the mechanism by which the liquid approaches the structure (and density) of DC, from which it readily crystallizes. We note that DC crystallization is avoided when the LL line drops below the glass transition temperature ($T_g$) of the HDL (~50 K above $T_o$ and $T_K$ by assuming $D(T_g)=10^{-20}$cm$^2$/s[20]), preventing the HDL to LDL transformation. A simple extrapolation from the data of Fig. 1 suggests that $T_{LL}=T_g^{HDL}$ for $\lambda\sim19.5$ ($\lambda=20$ in simulations, for which ergodicity is lost at high T). Our results point to the central role of LDL as an intermediate stage in the formation of the tetrahedral crystal.

We now discuss the fate of the LL equilibrium for $\lambda < 20$, inaccessible to MD simulations due to the glass transition of HDL. As HDL is the high temperature liquid, $S^{HDL} > S^{LDL}$ and thus we expect $T_K^{LDL} > T_K^{HDL}$. Figure 1 indicates that for $\lambda\sim19$ $T_{LL}=T_K^{HDL}$, making $\Delta S_{LL}$ very small (the difference in vibrational entropy of the two ideal glasses). The slope of the coexistence line, $dT/d\lambda \propto \Delta S_{LL}^{-1}$, will be almost vertical for $\lambda\sim19$. This suggests that LDL is stable only within the DC domain, $\lambda >18.75$.

This conclusion agrees with the expectation that the structure of the liquid will approach that of the stable crystal on cooling. Liquids such as silicon that seem to violate that rule at high temperature, eventually convert through a liquid-liquid transition to a more stable phase similar to the ground state crystal. In terms of classical nucleation theory, the similarity of structures between liquid and crystal at low temperature will cause the interfacial free energy between crystal nucleus and liquid to vanish so that nucleation becomes unavoidable, resolving the "Kauzmann paradox" in the way Kauzmann suggested.[21]

How do these glassformer models relate to any real materials? Our weakening of the tetrahedral bonding tendency is qualitatively what is achieved in the periodic table group IV by making the atomic core larger down the series. A measure of the strength of the tetrahedral interactions is the pressure that must be applied to transform DC into a higher coordination crystal. The pressure of the minimum temperature DC-βtin-liquid triple point decreases from 10.5 GPa for Si[22] through 9.5 GPa for Ge[23] to an extrapolated -0.5GPa for Sn[24]. Our mod-SW glassformer $\lambda=18.75$ would correspond to a hypothetical element intermediate between Ge and Sn for which the triple point occurs at p=0.

This letter shows that thermodynamics and glass formation are intrinsically related for mod-SW liquids. We propose that this relationship may be generalized to other liquids that present similar thermodynamic relations. Tetrahedral liquids, such as silicon[22], water and germanium[23] have T-p phase diagrams with the same minimum temperature triple point and polyamorphism mod-SW presents in T-$\lambda$. We expect them to present the same phenomenology discussed here. Tanaka has arrived to a similar conclusion based on his two-order-parameter model of liquids.[25] Recent indications of polyamorphous transformation in the path of ice crystallization from high density amorphous water[26] suggest that the role of the LDL as an intermediate in the tetrahedral crystal formation may be general to this class of liquids. We expect that water, silicon and germanium will have their optimum GFA at pressures around their low temperature triple point, probably where the LL and Kauzmann lines meet. Results of laboratory studies of rapid cooling of liquid Si and Ge at pressures close to their DC-β-tin-liquid triple points, and of MD simulations of SW Si under pressure, will be reported separately.[27]

For multi-phase systems with interfaces (e.g. colloidal suspensions exhibiting coexistence) the condition of glass formation at zero pressure that we attain with the mod-SW model is mandatory to explore the glass-gas region of the phase diagram. The use of multicomponent glassformers to explore systems with interfaces, that have been used to date for lack of better models, is particularly inadequate because of the surface segregation produced by the asymmetry of the interactions.[28]

This research was supported by NSF under Chemical Sciences grant no. 0404714.

*Corresponding author, email: valeria.molinero@utah.edu

**References:**

[1]     W. Klement, R. H. Willens, and P. Duwez, Nature **187**, 869 (1960).
[2]     J. Schroers, and W. L. Johnson, Appl Phys Lett **80**, 2069 (2002).
[3]     M. Dzugutov, Physical Review A **46**, R2984 (1992).
[4]     M. Dzugutov, Physical Review Letters **79**, 4043 (1997).
[5]     J. Roth, and A. R. Denton, Phys Rev E **61**, 6845 (2000).
[6]     W. Kob, and H. C. Andersen, Phys Rev E **51**, 4626 (1995).
[7]     A. Inoue, Acta Materalia **48**, 279 (2000).
[8]     F. H. Stillinger, and T. A. Weber, Phys Rev B **31**, 5262 (1985).
[9]     S. Sastry, and C. A. Angell, Nat Mater **2**, 739 (2003).
[10]    D. Turnbull, J Appl Phys **21**, 1022 (1950).
[11]    W. H. Wang, C. Dong, and C. H. Shek, Mat Sci Eng R **44**, 45 (2004).
[12]    The melting temperatures $T_m$ for each potential were obtained by stepwise heating (25K per 1.5 ns step) of DC and BCC ill-formed crystals obtained by freezing during a previous stepwise cooling (25 K each 3 ns) step, for λ=21.5 and 16, respectively. Parameter changes were implemented before each heating run. The melting, signaled by sudden energy absorption and loss of structure,

occurs sharply, in less than 10 ps. The method gives $T_m$ for DC ~80K below that determined by the phase coexistence method (M. Wilson, and P. F. McMillan, Phys. Rev. Lett. 90 (2003)]) and $T_m$ for BCC ~60K above the determination by the method of Yip and col., Phys Rev B 40, 2831 (1989) [Vitaliy Kapko, personal communication].

[13]     EPAPS, http://www.aip.org/pubservs/epaps.html  (Supporting Material).

[14]     R. Busch, Y. J. Kim, and W. L. Johnson, J Appl Phys **77**, 4039 (1995).

[15]     R. Bohmer, K. L. Ngai, C. A. Angell, and D. J. Plazek, Journal of Chemical Physics **99**, 4201 (1993).

[16]     S. M. Chathoth, A. Meyer, M. M. Koza, and F. Juranyi, Appl Phys Lett **85**, 4881 (2004).

[17]     G. Adam, and J. H. Gibbs, Journal of Chemical Physics **43**, 139 (1965).

[18]      An alternative definition of TK is based on the configurational entropy of the liquid $S_c$, $T_K=T(S_c=0)$.  The difference between $S^{ex}$ and $S_c$ is the excess vibrational entropy, that decreases with temperature as the liquid sample less shallow basin. Simulations of SPC/E water show the difference between the two definitions of $T_K$ to be 5K (A. Scala, F. W. Starr, E. La Nave, et al., Nature **406**, 166 (2000)) well below the uncertainty of our estimation.

[19]     S. Sastry, Physical Review Letters **85**, 590 (2000).

[20]     S. F. Swallen, P. A. Bonvallet, R. J. McMahon, and M. D. Ediger, Physical Review Letters **90**, 015901 (2003).

[21]     W. Kauzmann, Chemical Reviews **43**, 219 (1948).

[22]     G. A. Voronin, C. Pantea, T. W. Zerda, L. Wang, and Y. Zhao, Phys Rev B **68**, 020102 (2003).

[23]     G. A. Voronin, C. Pantea, T. W. Zerda, J. Zhang, L. Wang, and Y. Zhao, J Phys Chem Solids **64**, 2113 (2003).

[24]     J. Hafner, Phys Rev B **10**, 4151 (1974).

[25]     H. Tanaka, Phys Rev B **66**, 064202 (2002).

[26]     O. Mishima, Journal of Chemical Physics **123**, 154506 (2005).

[27]     H. Bhat, V. Molinero, J. Yarger, S. Sastry, and C. A. Angell, (in preparation).

[28]     A. S. Clarke, R. Kapral, and G. N. Patey, Journal of Chemical Physics **101**, 2432 (1994).

**Figure 1.** (color online) Phase diagram of the mod-SW potential. Melting lines for DC (down-pointing triangles) and BCC (squares) crystals, in relation to the "tetrahedrality" parameter λ, cross at the BCC-DC-liquid triple point $λ^{TP}$=18.75 and $T^{TP}$ =755 K. $λ^{TP}$ is very close to the T=0K  DC-BCC coexistence point, λ=18.6 (cross) obtained from equating the lattice energies. The dashed line

between these two points separates the DC and BCC domains. Lattice energies of several crystalline structures as a function of λ are shown in [13]. A β-tin phase, marginally stable in the range λ =18.2-18.7[13], is never seen. The glassforming domain, λ =17.5 to 20.25, is indicated by a bold line on λ axis. The temperature of zero mobility $T_o$ (blue triangles) and isoentropy Kauzmann temperature $T_K$ (red circles) coincide, within their error bars (±35 K for $T_o$ and ±12 K for $T_K$[13]). The minimum in $T_K$ occurs at the λ value where the isothermal diffusivity is the highest (Figure 3). Our melting lines end close to $T_K$, beyond which liquid equilibration is impossible. These endpoints have $\Delta H_m$~0 and $\Delta S_m$~0. The liquid-liquid transition temperatures, $T_{LL}$, are also shown in the range where the transition is observed in the simulations. A LL critical point is expected at high λ, but fast crystallization prevents the precise determination of $T_{LL}$ for λ > 21.5. To lower λ, we predict the LL will drop almost vertically after crossing $T_K$ (see text). The intersection of LL and HDL glass line (~50 K above $T_o$, see text) frustrates DC crystallization that occurs through the intermediate LDL phase.

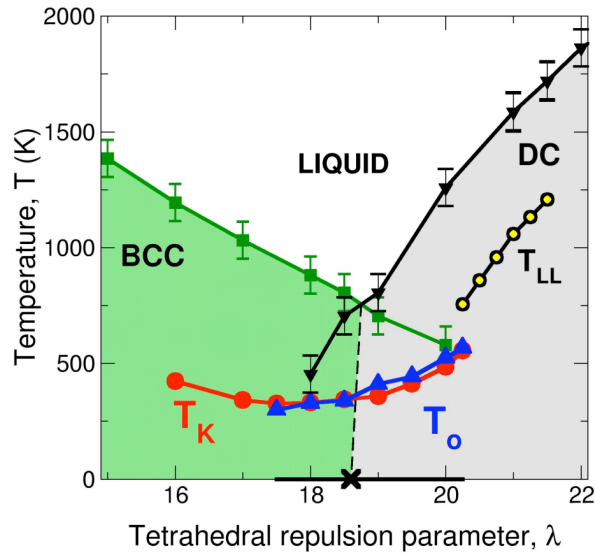

**Figure 2.** (color online) Excess thermodynamic functions of the liquid with respect to the stable perfect crystal (red: BCC, black: DC) for systems beyond the liquid-liquid transition region of the system. (a) Excess heat capacity $C_p^{ex}=C_p^{liquid}-C_p^{crystal}$ data for $\lambda < 20.25$, derived from fits of H vs T[13]. Symbols indicate the lowest temperature of equilibrated liquid in the simulation. (b) Decrease of excess entropy, $S^{ex}=S^{liquid}-S^{crystal}$ (Eq. 1), from its $\Delta S_m$ value (symbols) at defect crystal fusion point $T_m$. The condition $S^{ex}=0$ defines the Kauzmann temperatures, $T_K$. (c) Excess free energies $G^{ex}$ in the supercooled regime from $T_m$ to $T_K$. Symbols at the end of each $G^{ex}$ curve are only for labeling purposes. (d) $G^{ex}(T/T_m=0.8)$ rises asymmetrically on the two sides of the phase diagram (BCC in red, DC in black).

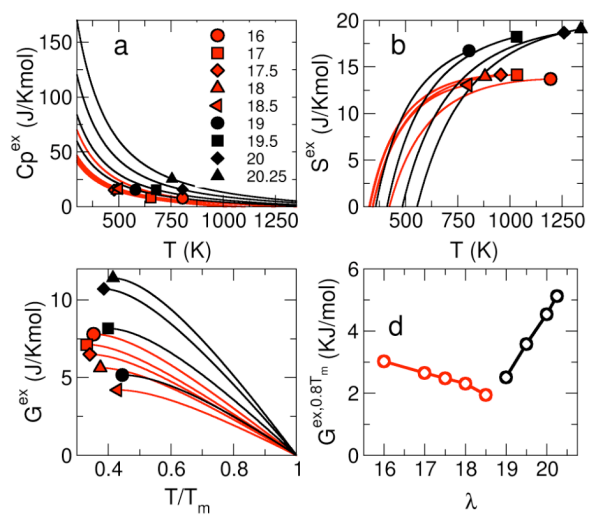

**Figure 3.** (color online) Arrhenius plots of diffusivities for non-crystallizing liquids of different $\lambda$ values. Diffusion coefficients D were obtained from linear fits of the MSD, $\langle r^2(t)\rangle=6Dt$. Solid curves are best fits of Eq. (2) to the data. The temperatures of zero mobility $T_o$ obtained from these D data are shown in Fig. 1., and coincide with the Kauzmann temperatures obtained from thermodynamics. The fragility parameter *F* of Eq.(2) for glassforming high density liquids is ~0.3 [13]. Insert shows the value of the liquid diffusivity at the melting points, $D(T_m)$, as a function of $\lambda$ (circles correspond to DC melting and triangles to BCC melting).

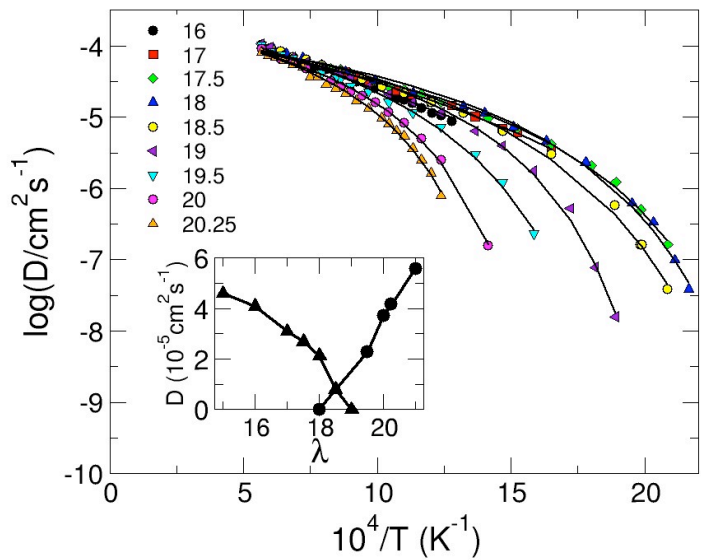